\documentclass[10pt,envcountsame,runningheads]{article}
\usepackage{amsmath,amsfonts,amssymb}
\makeatletter

\let\if@envcntreset\iffalse
\let\if@custvec\iftrue
\let\if@envcntsame\iffalse
\let\if@envcntsect\iffalse
\let\if@runhead\iffalse
\let\if@openbib\iffalse
\RequirePackage{multicol} 

\def\thisbottomragged{\def\@textbottom{\vskip\z@ plus.0001fil
\global\let\@textbottom\relax}}

\renewcommand\small{%
   \@setfontsize\small\@ixpt{11}%
   \abovedisplayskip 8.5\p@ \@plus3\p@ \@minus4\p@
   \abovedisplayshortskip \z@ \@plus2\p@
   \belowdisplayshortskip 4\p@ \@plus2\p@ \@minus2\p@
   \def\@listi{\leftmargin\leftmargini
               \parsep 0\p@ \@plus1\p@ \@minus\p@
               \topsep 8\p@ \@plus2\p@ \@minus4\p@
               \itemsep0\p@}%
   \belowdisplayskip \abovedisplayskip
}

\newcounter {chapter}
\renewcommand\thechapter      {\@arabic\c@chapter}

\newif\if@mainmatter \@mainmattertrue
\newcommand\frontmatter{\cleardoublepage
            \@mainmatterfalse\pagenumbering{Roman}}
\newcommand\mainmatter{\cleardoublepage
       \@mainmattertrue\pagenumbering{arabic}}
\newcommand\backmatter{\if@openright\cleardoublepage\else\clearpage\fi
      \@mainmatterfalse}

\renewcommand\part{\cleardoublepage
                 \thispagestyle{empty}%
                 \if@twocolumn
                     \onecolumn
                     \@tempswatrue
                   \else
                     \@tempswafalse
                 \fi
                 \null\vfil
                 \secdef\@part\@spart}

\def\@part[#1]#2{%
    \ifnum \c@secnumdepth >-2\relax
      \refstepcounter{part}%
      \addcontentsline{toc}{part}{\thepart\hspace{1em}#1}%
    \else
      \addcontentsline{toc}{part}{#1}%
    \fi
    \markboth{}{}%
    {\centering
     \interlinepenalty \@M
     \normalfont
     \ifnum \c@secnumdepth >-2\relax
       \huge\bfseries \partname~\thepart
       \par
       \vskip 20\p@
     \fi
     \Huge \bfseries #2\par}%
    \@endpart}
\def\@spart#1{%
    {\centering
     \interlinepenalty \@M
     \normalfont
     \Huge \bfseries #1\par}%
    \@endpart}
\def\@endpart{\vfil\newpage
              \if@twoside
                \null
                \thispagestyle{empty}%
                \newpage
              \fi
              \if@tempswa
                \twocolumn
              \fi}

\newcommand\chapter{\clearpage
                    \thispagestyle{empty}%
                    \global\@topnum\z@
                    \@afterindentfalse
                    \secdef\@chapter\@schapter}
\def\@chapter[#1]#2{\ifnum \c@secnumdepth >\m@ne
                       \if@mainmatter
                         \refstepcounter{chapter}%
                         \typeout{\@chapapp\space\thechapter.}%
                         \addcontentsline{toc}{chapter}%
                                  {\protect\numberline{\thechapter}#1}%
                       \else
                         \addcontentsline{toc}{chapter}{#1}%
                       \fi
                    \else
                      \addcontentsline{toc}{chapter}{#1}%
                    \fi
                    \chaptermark{#1}%
                    \addtocontents{lof}{\protect\addvspace{10\p@}}%
                    \addtocontents{lot}{\protect\addvspace{10\p@}}%
                    \if@twocolumn
                      \@topnewpage[\@makechapterhead{#2}]%
                    \else
                      \@makechapterhead{#2}%
                      \@afterheading
                    \fi}
\def\@makechapterhead#1{%
  {\centering
    \ifnum \c@secnumdepth >\m@ne
      \if@mainmatter
        \large\bfseries \@chapapp{} \thechapter
        \par\nobreak
        \vskip 20\p@
      \fi
    \fi
    \interlinepenalty\@M
    \Large \bfseries #1\par\nobreak
    \vskip 40\p@
  }}
\def\@schapter#1{\if@twocolumn
                   \@topnewpage[\@makeschapterhead{#1}]%
                 \else
                   \@makeschapterhead{#1}%
                   \@afterheading
                 \fi}
\def\@makeschapterhead#1{%
  {\centering
    \normalfont
    \interlinepenalty\@M
    \Large \bfseries  #1\par\nobreak
    \vskip 40\p@
  }}

\renewcommand\section{\@startsection{section}{1}{\z@}%
                       {-18\p@ \@plus -4\p@ \@minus -4\p@}%
                       {12\p@ \@plus 4\p@ \@minus 4\p@}%
                       {\normalfont\large\bfseries\boldmath
                        \rightskip=\z@ \@plus 8em\pretolerance=10000 }}
\renewcommand\subsection{\@startsection{subsection}{2}{\z@}%
                       {-18\p@ \@plus -4\p@ \@minus -4\p@}%
                       {8\p@ \@plus 4\p@ \@minus 4\p@}%
                       {\normalfont\normalsize\bfseries\boldmath
                        \rightskip=\z@ \@plus 8em\pretolerance=10000 }}
\renewcommand\subsubsection{\@startsection{subsubsection}{3}{\z@}%
                       {-18\p@ \@plus -4\p@ \@minus -4\p@}%
                       {-0.5em \@plus -0.22em \@minus -0.1em}%
                       {\normalfont\normalsize\bfseries\boldmath}}
\renewcommand\paragraph{\@startsection{paragraph}{4}{\z@}%
                       {-12\p@ \@plus -4\p@ \@minus -4\p@}%
                       {-0.5em \@plus -0.22em \@minus -0.1em}%
                       {\normalfont\normalsize\itshape}}
\renewcommand\subparagraph[1]{\typeout{LLNCS warning: You should not use
                  \string\subparagraph\space with this class}\vskip0.5cm
You should not use \verb|\subparagraph| with this class.\vskip0.5cm}

\DeclareMathSymbol{\Gamma}{\mathalpha}{letters}{"00}
\DeclareMathSymbol{\Delta}{\mathalpha}{letters}{"01}
\DeclareMathSymbol{\Theta}{\mathalpha}{letters}{"02}
\DeclareMathSymbol{\Lambda}{\mathalpha}{letters}{"03}
\DeclareMathSymbol{\Xi}{\mathalpha}{letters}{"04}
\DeclareMathSymbol{\Pi}{\mathalpha}{letters}{"05}
\DeclareMathSymbol{\Sigma}{\mathalpha}{letters}{"06}
\DeclareMathSymbol{\Upsilon}{\mathalpha}{letters}{"07}
\DeclareMathSymbol{\Phi}{\mathalpha}{letters}{"08}
\DeclareMathSymbol{\Psi}{\mathalpha}{letters}{"09}
\DeclareMathSymbol{\Omega}{\mathalpha}{letters}{"0A}

\if@custvec

\fi

\def\squareforqed{\hbox{\rlap{$\sqcap$}$\sqcup$}}
\def\qed{\ifmmode\squareforqed\else{\unskip\nobreak\hfil
\penalty50\hskip1em\null\nobreak\hfil\squareforqed
\parfillskip=0pt\finalhyphendemerits=0\endgraf}\fi}

\def\bbbc{{\mathchoice {\setbox0=\hbox{$\displaystyle\rm C$}\hbox{\hbox
to0pt{\kern0.4\wd0\vrule height0.9\ht0\hss}\box0}}
{\setbox0=\hbox{$\textstyle\rm C$}\hbox{\hbox
to0pt{\kern0.4\wd0\vrule height0.9\ht0\hss}\box0}}
{\setbox0=\hbox{$\scriptstyle\rm C$}\hbox{\hbox
to0pt{\kern0.4\wd0\vrule height0.9\ht0\hss}\box0}}
{\setbox0=\hbox{$\scriptscriptstyle\rm C$}\hbox{\hbox
to0pt{\kern0.4\wd0\vrule height0.9\ht0\hss}\box0}}}}
\def\bbbq{{\mathchoice {\setbox0=\hbox{$\displaystyle\rm
Q$}\hbox{\raise
0.15\ht0\hbox to0pt{\kern0.4\wd0\vrule height0.8\ht0\hss}\box0}}
{\setbox0=\hbox{$\textstyle\rm Q$}\hbox{\raise
0.15\ht0\hbox to0pt{\kern0.4\wd0\vrule height0.8\ht0\hss}\box0}}
{\setbox0=\hbox{$\scriptstyle\rm Q$}\hbox{\raise
0.15\ht0\hbox to0pt{\kern0.4\wd0\vrule height0.7\ht0\hss}\box0}}
{\setbox0=\hbox{$\scriptscriptstyle\rm Q$}\hbox{\raise
0.15\ht0\hbox to0pt{\kern0.4\wd0\vrule height0.7\ht0\hss}\box0}}}}
\def\bbbt{{\mathchoice {\setbox0=\hbox{$\displaystyle\rm
T$}\hbox{\hbox to0pt{\kern0.3\wd0\vrule height0.9\ht0\hss}\box0}}
{\setbox0=\hbox{$\textstyle\rm T$}\hbox{\hbox
to0pt{\kern0.3\wd0\vrule height0.9\ht0\hss}\box0}}
{\setbox0=\hbox{$\scriptstyle\rm T$}\hbox{\hbox
to0pt{\kern0.3\wd0\vrule height0.9\ht0\hss}\box0}}
{\setbox0=\hbox{$\scriptscriptstyle\rm T$}\hbox{\hbox
to0pt{\kern0.3\wd0\vrule height0.9\ht0\hss}\box0}}}}
\def\bbbs{{\mathchoice
{\setbox0=\hbox{$\displaystyle     \rm S$}\hbox{\raise0.5\ht0\hbox
to0pt{\kern0.35\wd0\vrule height0.45\ht0\hss}\hbox
to0pt{\kern0.55\wd0\vrule height0.5\ht0\hss}\box0}}
{\setbox0=\hbox{$\textstyle        \rm S$}\hbox{\raise0.5\ht0\hbox
to0pt{\kern0.35\wd0\vrule height0.45\ht0\hss}\hbox
to0pt{\kern0.55\wd0\vrule height0.5\ht0\hss}\box0}}
{\setbox0=\hbox{$\scriptstyle      \rm S$}\hbox{\raise0.5\ht0\hbox
to0pt{\kern0.35\wd0\vrule height0.45\ht0\hss}\raise0.05\ht0\hbox
to0pt{\kern0.5\wd0\vrule height0.45\ht0\hss}\box0}}
{\setbox0=\hbox{$\scriptscriptstyle\rm S$}\hbox{\raise0.5\ht0\hbox
to0pt{\kern0.4\wd0\vrule height0.45\ht0\hss}\raise0.05\ht0\hbox
to0pt{\kern0.55\wd0\vrule height0.45\ht0\hss}\box0}}}}
\def\bbbz{{\mathchoice {\hbox{$\mathsf\textstyle Z\kern-0.4em Z$}}
{\hbox{$\mathsf\textstyle Z\kern-0.4em Z$}}
{\hbox{$\mathsf\scriptstyle Z\kern-0.3em Z$}}
{\hbox{$\mathsf\scriptscriptstyle Z\kern-0.2em Z$}}}}

\let\ts\,

\def\@listI{\leftmargin\leftmargini
            \parsep 0\p@ \@plus1\p@ \@minus\p@
            \topsep 8\p@ \@plus2\p@ \@minus4\p@
            \itemsep0\p@}
\let\@listi\@listI
\@listi
\def\@listii {\leftmargin\leftmarginii
              \labelwidth\leftmarginii
              \advance\labelwidth-\labelsep
              \topsep    0\p@ \@plus2\p@ \@minus\p@}
\def\@listiii{\leftmargin\leftmarginiii
              \labelwidth\leftmarginiii
              \advance\labelwidth-\labelsep
              \topsep    0\p@ \@plus\p@\@minus\p@
              \parsep    \z@
              \partopsep \p@ \@plus\z@ \@minus\p@}

\renewcommand\labelitemii{$\m@th\bullet$}

\def\tableofcontents{\chapter*{\contentsname\@mkboth{{\contentsname}}%
                                                    {{\contentsname}}}
 \def\authcount##1{\setcounter{auco}{##1}\setcounter{@auth}{1}}
 \def\lastand{\ifnum\value{auco}=2\relax
                 \unskip{} \andname\
              \else
                 \unskip \lastandname\
              \fi}%
 \def\and{\stepcounter{@auth}\relax
          \ifnum\value{@auth}=\value{auco}%
             \lastand
          \else
             \unskip,
          \fi}%
 \@starttoc{toc}\if@restonecol\twocolumn\fi}

\def\l@part#1#2{\addpenalty{\@secpenalty}%
   \addvspace{2em plus\p@}
   \begingroup
     \parindent \z@
     \rightskip \z@ plus 5em
     \hrule\vskip5pt
     \large               
     \bfseries\boldmath   
     \leavevmode          
     #1\par
     \vskip5pt
     \hrule
     \vskip1pt
     \nobreak             
   \endgroup}

\def\@dotsep{2}

\def\hyperhrefextend{\ifx\hyper@anchor\@undefined\else
{chapter.\thechapter}\fi}

\def\addnumcontentsmark#1#2#3{%
\addtocontents{#1}{\protect\contentsline{#2}{\protect\numberline
                     {\thechapter}#3}{\thepage}\hyperhrefextend}}
\def\addcontentsmark#1#2#3{%
\addtocontents{#1}{\protect\contentsline{#2}{#3}{\thepage}\hyperhrefextend}}
\def\addcontentsmarkwop#1#2#3{%
\addtocontents{#1}{\protect\contentsline{#2}{#3}{0}\hyperhrefextend}}

\def\@adcmk[#1]{\ifcase #1 \or
\def\@gtempa{\addnumcontentsmark}%
  \or    \def\@gtempa{\addcontentsmark}%
  \or    \def\@gtempa{\addcontentsmarkwop}%
  \fi\@gtempa{toc}{chapter}}
\def\addtocmark{\@ifnextchar[{\@adcmk}{\@adcmk[3]}}

\def\l@chapter#1#2{\addpenalty{-\@highpenalty}
 \vskip 1.0em plus 1pt \@tempdima 1.5em \begingroup
 \parindent \z@ \rightskip \@pnumwidth
 \parfillskip -\@pnumwidth
 \leavevmode \advance\leftskip\@tempdima \hskip -\leftskip
 {\large\bfseries\boldmath#1}\ifx0#2\hfil\null
 \else
      \nobreak
      \leaders\hbox{$\m@th \mkern \@dotsep mu.\mkern
      \@dotsep mu$}\hfill
      \nobreak\hbox to\@pnumwidth{\hss #2}%
 \fi\par
 \penalty\@highpenalty \endgroup}

\def\l@title#1#2{\addpenalty{-\@highpenalty}
 \addvspace{8pt plus 1pt}
 \@tempdima \z@
 \begingroup
 \parindent \z@ \rightskip \@tocrmarg
 \parfillskip -\@tocrmarg
 \leavevmode \advance\leftskip\@tempdima \hskip -\leftskip
 #1\nobreak
 \leaders\hbox{$\m@th \mkern \@dotsep mu.\mkern
 \@dotsep mu$}\hfill
 \nobreak\hbox to\@pnumwidth{\hss #2}\par
 \penalty\@highpenalty \endgroup}

\newdimen\tocchpnum
\newdimen\tocsecnum
\newdimen\tocsectotal
\newdimen\tocsubsecnum
\newdimen\tocsubsectotal
\newdimen\tocsubsubsecnum
\newdimen\tocsubsubsectotal
\newdimen\tocparanum
\newdimen\tocparatotal
\newdimen\tocsubparanum
\tocchpnum=\z@            
\def\calctocindent{%
\tocsectotal=\tocchpnum
\advance\tocsectotal by\tocsecnum
\tocsubsectotal=\tocsectotal
\advance\tocsubsectotal by\tocsubsecnum
\tocsubsubsectotal=\tocsubsectotal
\advance\tocsubsubsectotal by\tocsubsubsecnum
\tocparatotal=\tocsubsubsectotal
\advance\tocparatotal by\tocparanum}
\calctocindent

\def\l@section{\@dottedtocline{1}{\tocchpnum}{\tocsecnum}}
\def\l@subsection{\@dottedtocline{2}{\tocsectotal}{\tocsubsecnum}}
\def\l@subsubsection{\@dottedtocline{3}{\tocsubsectotal}{\tocsubsubsecnum}}
\def\l@paragraph{\@dottedtocline{4}{\tocsubsubsectotal}{\tocparanum}}
\def\l@subparagraph{\@dottedtocline{5}{\tocparatotal}{\tocsubparanum}}

\def\listoffigures{\@restonecolfalse\if@twocolumn\@restonecoltrue\onecolumn
 \fi\section*{\listfigurename\@mkboth{{\listfigurename}}{{\listfigurename}}}
 \@starttoc{lof}\if@restonecol\twocolumn\fi}
\def\l@figure{\@dottedtocline{1}{0em}{1.5em}}

\def\listoftables{\@restonecolfalse\if@twocolumn\@restonecoltrue\onecolumn
 \fi\section*{\listtablename\@mkboth{{\listtablename}}{{\listtablename}}}
 \@starttoc{lot}\if@restonecol\twocolumn\fi}
\let\l@table\l@figure

\renewcommand\listoffigures{%
    \section*{\listfigurename
      \@mkboth{\listfigurename}{\listfigurename}}%
    \@starttoc{lof}%
    }

\renewcommand\listoftables{%
    \section*{\listtablename
      \@mkboth{\listtablename}{\listtablename}}%
    \@starttoc{lot}%
    }

\ifx\oribibl\undefined
\ifx\citeauthoryear\undefined
\renewenvironment{thebibliography}[1]
     {\section*{\refname}
      \def\@biblabel##1{##1.}
      \small
      \list{\@biblabel{\@arabic\c@enumiv}}%
           {\settowidth\labelwidth{\@biblabel{#1}}%
            \leftmargin\labelwidth
            \advance\leftmargin\labelsep
            \if@openbib
              \advance\leftmargin\bibindent
              \itemindent -\bibindent
              \listparindent \itemindent
              \parsep \z@
            \fi
            \usecounter{enumiv}%
            \let\p@enumiv\@empty
            \renewcommand\theenumiv{\@arabic\c@enumiv}}%
      \if@openbib
        \renewcommand\newblock{\par}%
      \else
        \renewcommand\newblock{\hskip .11em \@plus.33em \@minus.07em}%
      \fi
      \sloppy\clubpenalty4000\widowpenalty4000%
      \sfcode`\.=\@m}
     {\def\@noitemerr
       {\@latex@warning{Empty `thebibliography' environment}}%
      \endlist}
\def\@lbibitem[#1]#2{\item[{[#1]}\hfill]\if@filesw
     {\let\protect\noexpand\immediate
     \write\@auxout{\string\bibcite{#2}{#1}}}\fi\ignorespaces}
\newcount\@tempcntc
\def\@citex[#1]#2{\if@filesw\immediate\write\@auxout{\string\citation{#2}}\fi
  \@tempcnta\z@\@tempcntb\m@ne\def\@citea{}\@cite{\@for\@citeb:=#2\do
    {\@ifundefined
       {b@\@citeb}{\@citeo\@tempcntb\m@ne\@citea\def\@citea{,}{\bfseries
        ?}\@warning
       {Citation `\@citeb' on page \thepage \space undefined}}%
    {\setbox\z@\hbox{\global\@tempcntc0\csname b@\@citeb\endcsname\relax}%
     \ifnum\@tempcntc=\z@ \@citeo\@tempcntb\m@ne
       \@citea\def\@citea{,}\hbox{\csname b@\@citeb\endcsname}%
     \else
      \advance\@tempcntb\@ne
      \ifnum\@tempcntb=\@tempcntc
      \else\advance\@tempcntb\m@ne\@citeo
      \@tempcnta\@tempcntc\@tempcntb\@tempcntc\fi\fi}}\@citeo}{#1}}
\def\@citeo{\ifnum\@tempcnta>\@tempcntb\else
               \@citea\def\@citea{,\,\hskip\z@skip}%
               \ifnum\@tempcnta=\@tempcntb\the\@tempcnta\else
               {\advance\@tempcnta\@ne\ifnum\@tempcnta=\@tempcntb \else
                \def\@citea{--}\fi
      \advance\@tempcnta\m@ne\the\@tempcnta\@citea\the\@tempcntb}\fi\fi}
\else
\renewenvironment{thebibliography}[1]
     {\section*{\refname}
      \small
      \list{}%
           {\settowidth\labelwidth{}%
            \leftmargin\parindent
            \itemindent=-\parindent
            \labelsep=\z@
            \if@openbib
              \advance\leftmargin\bibindent
              \itemindent -\bibindent
              \listparindent \itemindent
              \parsep \z@
            \fi
            \usecounter{enumiv}%
            \let\p@enumiv\@empty
            \renewcommand\theenumiv{}}%
      \if@openbib
        \renewcommand\newblock{\par}%
      \else
        \renewcommand\newblock{\hskip .11em \@plus.33em \@minus.07em}%
      \fi
      \sloppy\clubpenalty4000\widowpenalty4000%
      \sfcode`\.=\@m}
     {\def\@noitemerr
       {\@latex@warning{Empty `thebibliography' environment}}%
      \endlist}
      \def\@cite#1{#1}%
      \def\@lbibitem[#1]#2{\item[]\if@filesw
        {\def\protect##1{\string ##1\space}\immediate
      \write\@auxout{\string\bibcite{#2}{#1}}}\fi\ignorespaces}
   \fi
\else
\@cons\@openbib@code{\noexpand\small}
\fi

\def\idxquad{\hskip 10\p@}

\def\@idxitem{\par\hangindent 10\p@}

\def\indexspace{\par \vskip 10\p@ plus5\p@ minus3\p@\relax}

\renewcommand\footnoterule{%
  \kern-3\p@
  \hrule\@width 2truecm
  \kern2.6\p@}
  \newdimen\fnindent
\long\def\@makefntext#1{%
    \parindent \fnindent%
    \leftskip \fnindent%
    \noindent
    \llap{\hb@xt@1em{\hss\@makefnmark\ }}\ignorespaces#1}

\long\def\@makecaption#1#2{%
  \vskip\abovecaptionskip
  \sbox\@tempboxa{{\bfseries #1.} #2}%
  \ifdim \wd\@tempboxa >\hsize
    {\bfseries #1.} #2\par
  \else
    \global \@minipagefalse
    \hb@xt@\hsize{\hfil\box\@tempboxa\hfil}%
  \fi
  \vskip\belowcaptionskip}

\def\fps@figure{htbp}
\def\fnum@figure{\figurename\thinspace\thefigure}
\def \@floatboxreset {%
        \reset@font
        \small
        \@setnobreak
        \@setminipage
}
\def\fps@table{htbp}
\def\fnum@table{\tablename~\thetable}

\renewenvironment{table*}
               {\setlength\abovecaptionskip{0\p@}%
                \setlength\belowcaptionskip{10\p@}%
                \@dblfloat{table}}
               {\end@dblfloat}

\long\def\@caption#1[#2]#3{\par\addcontentsline{\csname
  ext@#1\endcsname}{#1}{\protect\numberline{\csname
  the#1\endcsname}{\ignorespaces #2}}\begingroup
    \@parboxrestore
    \@makecaption{\csname fnum@#1\endcsname}{\ignorespaces #3}\par
  \endgroup}

\newcounter{@inst}
\newcounter{@auth}
\newcounter{auco}
\def\andname{and}
\def\lastandname{\unskip, and}
\newdimen\instindent
\newbox\authrun
\newtoks\authorrunning
\newtoks\tocauthor
\newbox\titrun
\newtoks\titlerunning
\newtoks\toctitle

\def\clearheadinfo{\gdef\@author{No Author Given}%
                   \gdef\@title{No Title Given}%
                   \gdef\@subtitle{}%
                   \gdef\@institute{No Institute Given}%
                   \gdef\@thanks{}%
                   \global\titlerunning={}\global\authorrunning={}%
                   \global\toctitle={}\global\tocauthor={}}

\def\institute#1{\gdef\@institute{#1}}

\def\institutename{\par
 \begingroup
 \parskip=\z@
 \parindent=\z@
 \setcounter{@inst}{1}%
 \def\and{\par\stepcounter{@inst}%
 \noindent$^{\the@inst}$\enspace\ignorespaces}%
 \setbox0=\vbox{\def\thanks##1{}\@institute}%
 \ifnum\c@@inst=1\relax
 \else
   \setcounter{footnote}{\c@@inst}%
   \setcounter{@inst}{1}%
   \noindent$^{\the@inst}$\enspace
 \fi
 \ignorespaces
 \@institute\par
 \endgroup}

\def\@fnsymbol#1{\ensuremath{\ifcase#1\or\star\or{\star\star}\or
   {\star\star\star}\or \dagger\or \ddagger\or
   \mathchar "278\or \mathchar "27B\or \|\or **\or \dagger\dagger
   \or \ddagger\ddagger \else\@ctrerr\fi}}

\def\inst#1{\unskip$^{#1}$}
\def\fnmsep{\unskip$^,$}
\def\email#1{{\tt#1}}
\AtBeginDocument{\@ifundefined{url}{\def\url#1{#1}}{}}

\def\subtitle#1{\gdef\@subtitle{#1}}
\clearheadinfo

\renewcommand\maketitle{\newpage
  \refstepcounter{chapter}%
  \stepcounter{section}%
  \setcounter{section}{0}%
  \setcounter{subsection}{0}%
  \setcounter{figure}{0}
  \setcounter{table}{0}
  \setcounter{equation}{0}
  \setcounter{footnote}{0}%
  \begingroup
    \parindent=\z@
    \renewcommand\thefootnote{\@fnsymbol\c@footnote}%
    \if@twocolumn
      \ifnum \col@number=\@ne
        \@maketitle
      \else
        \twocolumn[\@maketitle]%
      \fi
    \else
      \newpage
      \global\@topnum\z@   
      \@maketitle
    \fi
    \thispagestyle{empty}\@thanks
    \def\\{\unskip\ \ignorespaces}\def\inst##1{\unskip{}}%
    \def\thanks##1{\unskip{}}\def\fnmsep{\unskip}%
    \instindent=\hsize
    \advance\instindent by-\headlineindent
    \if!\the\toctitle!\addcontentsline{toc}{title}{\@title}\else
       \addcontentsline{toc}{title}{\the\toctitle}\fi
    \if@runhead
       \if!\the\titlerunning!\else
         \edef\@title{\the\titlerunning}%
       \fi
       \global\setbox\titrun=\hbox{\small\rm\unboldmath\ignorespaces\@title}%
       \ifdim\wd\titrun>\instindent
          \typeout{Title too long for running head. Please supply}%
          \typeout{a shorter form with \string\titlerunning\space prior to
                   \string\maketitle}%
          \global\setbox\titrun=\hbox{\small\rm
          Title Suppressed Due to Excessive Length}%
       \fi
       \xdef\@title{\copy\titrun}%
    \fi
    \if!\the\tocauthor!\relax
      {\def\and{\noexpand\protect\noexpand\and}%
      \protected@xdef\toc@uthor{\@author}}%
    \else
      \def\\{\noexpand\protect\noexpand\newline}%
      \protected@xdef\scratch{\the\tocauthor}%
      \protected@xdef\toc@uthor{\scratch}%
    \fi
    \addtocontents{toc}{{\protect\raggedright\protect\leftskip15\p@
    \protect\rightskip\@tocrmarg
    \protect\itshape\toc@uthor\protect\endgraf}}%
    \if@runhead
       \if!\the\authorrunning!
         \value{@inst}=\value{@auth}%
         \setcounter{@auth}{1}%
       \else
         \edef\@author{\the\authorrunning}%
       \fi
       \global\setbox\authrun=\hbox{\small\unboldmath\@author\unskip}%
       \ifdim\wd\authrun>\instindent
          \typeout{Names of authors too long for running head. Please supply}%
          \typeout{a shorter form with \string\authorrunning\space prior to
                   \string\maketitle}%
          \global\setbox\authrun=\hbox{\small\rm
          Authors Suppressed Due to Excessive Length}%
       \fi
       \xdef\@author{\copy\authrun}%
       \markboth{\@author}{\@title}%
     \fi
  \endgroup
  \setcounter{footnote}{0}%
  \clearheadinfo}
\def\@maketitle{\newpage
 \markboth{}{}%
 \def\lastand{\ifnum\value{@inst}=2\relax
                 \unskip{} \andname\
              \else
                 \unskip \lastandname\
              \fi}%
 \def\and{\stepcounter{@auth}\relax
          \ifnum\value{@auth}=\value{@inst}%
             \lastand
          \else
             \unskip,
          \fi}%
 \begin{center}%
 {\Large \bfseries\boldmath
  \pretolerance=10000
  \@title \par}\vskip .8cm
\if!\@subtitle!\else {\large \bfseries\boldmath
  \vskip -.65cm
  \pretolerance=10000
  \@subtitle \par}\vskip .8cm\fi
 \setbox0=\vbox{\setcounter{@auth}{1}\def\and{\stepcounter{@auth}}%
 \def\thanks##1{}\@author}%
 \global\value{@inst}=\value{@auth}%
 \global\value{auco}=\value{@auth}%
 \setcounter{@auth}{1}%
{\lineskip .5em
\noindent\ignorespaces
\@author\vskip.35cm}
 {\small\institutename}
 \end{center}%
 }

\def\@thmcountersep{}
\def\@thmcounterend{.}

\def\spnewtheorem{\@ifstar{\@sthm}{\@Sthm}}

\def\@spnthm#1#2{%
  \@ifnextchar[{\@spxnthm{#1}{#2}}{\@spynthm{#1}{#2}}}
\def\@Sthm#1{\@ifnextchar[{\@spothm{#1}}{\@spnthm{#1}}}

\def\@spxnthm#1#2[#3]#4#5{\expandafter\@ifdefinable\csname #1\endcsname
   {\@definecounter{#1}\@addtoreset{#1}{#3}%
   \expandafter\xdef\csname the#1\endcsname{\expandafter\noexpand
     \csname the#3\endcsname \noexpand\@thmcountersep \@thmcounter{#1}}%
   \expandafter\xdef\csname #1name\endcsname{#2}%
   \global\@namedef{#1}{\@spthm{#1}{\csname #1name\endcsname}{#4}{#5}}%
                              \global\@namedef{end#1}{\@endtheorem}}}

\def\@spynthm#1#2#3#4{\expandafter\@ifdefinable\csname #1\endcsname
   {\@definecounter{#1}%
   \expandafter\xdef\csname the#1\endcsname{\@thmcounter{#1}}%
   \expandafter\xdef\csname #1name\endcsname{#2}%
   \global\@namedef{#1}{\@spthm{#1}{\csname #1name\endcsname}{#3}{#4}}%
                               \global\@namedef{end#1}{\@endtheorem}}}

\def\@spothm#1[#2]#3#4#5{%
  \@ifundefined{c@#2}{\@latexerr{No theorem environment `#2' defined}\@eha}%
  {\expandafter\@ifdefinable\csname #1\endcsname
  {\global\@namedef{the#1}{\@nameuse{the#2}}%
  \expandafter\xdef\csname #1name\endcsname{#3}%
  \global\@namedef{#1}{\@spthm{#2}{\csname #1name\endcsname}{#4}{#5}}%
  \global\@namedef{end#1}{\@endtheorem}}}}

\def\@spthm#1#2#3#4{\topsep 7\p@ \@plus2\p@ \@minus4\p@
\refstepcounter{#1}%
\@ifnextchar[{\@spythm{#1}{#2}{#3}{#4}}{\@spxthm{#1}{#2}{#3}{#4}}}

\def\@spxthm#1#2#3#4{\@spbegintheorem{#2}{\csname the#1\endcsname}{#3}{#4}%
                    \ignorespaces}

\def\@spythm#1#2#3#4[#5]{\@spopargbegintheorem{#2}{\csname
       the#1\endcsname}{#5}{#3}{#4}\ignorespaces}

\def\@spbegintheorem#1#2#3#4{\trivlist
                 \item[\hskip\labelsep{#3#1\ #2\@thmcounterend}]#4}

\def\@spopargbegintheorem#1#2#3#4#5{\trivlist
      \item[\hskip\labelsep{#4#1\ #2}]{#4(#3)\@thmcounterend\ }#5}

\def\@sthm#1#2{\@Ynthm{#1}{#2}}

\def\@Ynthm#1#2#3#4{\expandafter\@ifdefinable\csname #1\endcsname
   {\global\@namedef{#1}{\@Thm{\csname #1name\endcsname}{#3}{#4}}%
    \expandafter\xdef\csname #1name\endcsname{#2}%
    \global\@namedef{end#1}{\@endtheorem}}}

\def\@Thm#1#2#3{\topsep 7\p@ \@plus2\p@ \@minus4\p@
\@ifnextchar[{\@Ythm{#1}{#2}{#3}}{\@Xthm{#1}{#2}{#3}}}

\def\@Xthm#1#2#3{\@Begintheorem{#1}{#2}{#3}\ignorespaces}

\def\@Ythm#1#2#3[#4]{\@Opargbegintheorem{#1}
       {#4}{#2}{#3}\ignorespaces}

\def\@Begintheorem#1#2#3{#3\trivlist
                           \item[\hskip\labelsep{#2#1\@thmcounterend}]}

\def\@Opargbegintheorem#1#2#3#4{#4\trivlist
      \item[\hskip\labelsep{#3#1}]{#3(#2)\@thmcounterend\ }}

\if@envcntsect
   \def\@thmcountersep{.}
   \spnewtheorem{theorem}{Theorem}[section]{\bfseries}{\itshape}
\else
   \spnewtheorem{theorem}{Theorem}{\bfseries}{\itshape}
   \if@envcntreset
      \@addtoreset{theorem}{section}
   \else
      \@addtoreset{theorem}{chapter}
   \fi
\fi

\spnewtheorem*{claim}{Claim}{\itshape}{\rmfamily}
\spnewtheorem*{proof}{Proof}{\itshape}{\rmfamily}
\if@envcntsame 
   \def\spn@wtheorem#1#2#3#4{\@spothm{#1}[theorem]{#2}{#3}{#4}}
\else 
   \if@envcntsect 
      \def\spn@wtheorem#1#2#3#4{\@spxnthm{#1}{#2}[section]{#3}{#4}}
   \else 
      \if@envcntreset
         \def\spn@wtheorem#1#2#3#4{\@spynthm{#1}{#2}{#3}{#4}
                                   \@addtoreset{#1}{section}}
      \else
         \def\spn@wtheorem#1#2#3#4{\@spynthm{#1}{#2}{#3}{#4}
                                   \@addtoreset{#1}{chapter}}%
      \fi
   \fi
\fi
\spn@wtheorem{case}{Case}{\itshape}{\rmfamily}
\spn@wtheorem{conjecture}{Conjecture}{\itshape}{\rmfamily}
\spn@wtheorem{corollary}{Corollary}{\bfseries}{\itshape}
\spn@wtheorem{definition}{Definition}{\bfseries}{\itshape}
\spn@wtheorem{example}{Example}{\itshape}{\rmfamily}
\spn@wtheorem{exercise}{Exercise}{\itshape}{\rmfamily}
\spn@wtheorem{lemma}{Lemma}{\bfseries}{\itshape}
\spn@wtheorem{note}{Note}{\itshape}{\rmfamily}
\spn@wtheorem{problem}{Problem}{\itshape}{\rmfamily}
\spn@wtheorem{property}{Property}{\itshape}{\rmfamily}
\spn@wtheorem{proposition}{Proposition}{\bfseries}{\itshape}
\spn@wtheorem{question}{Question}{\itshape}{\rmfamily}
\spn@wtheorem{solution}{Solution}{\itshape}{\rmfamily}
\spn@wtheorem{remark}{Remark}{\itshape}{\rmfamily}

\def\@takefromreset#1#2{%
    \def\@tempa{#1}%
    \let\@tempd\@elt
    \def\@elt##1{%
        \def\@tempb{##1}%
        \ifx\@tempa\@tempb\else
            \@addtoreset{##1}{#2}%
        \fi}%
    \expandafter\expandafter\let\expandafter\@tempc\csname cl@#2\endcsname
    \expandafter\def\csname cl@#2\endcsname{}%
    \@tempc
    \let\@elt\@tempd}

\def\theopargself{\def\@spopargbegintheorem##1##2##3##4##5{\trivlist
      \item[\hskip\labelsep{##4##1\ ##2}]{##4##3\@thmcounterend\ }##5}
                  \def\@Opargbegintheorem##1##2##3##4{##4\trivlist
      \item[\hskip\labelsep{##3##1}]{##3##2\@thmcounterend\ }}
      }

\renewenvironment{abstract}{%
      \list{}{\advance\topsep by0.35cm\relax\small
      \leftmargin=1cm
      \labelwidth=\z@
      \listparindent=\z@
      \itemindent\listparindent
      \rightmargin\leftmargin}\item[\hskip\labelsep
                                    \bfseries\abstractname]}
    {\endlist}
\renewcommand{\abstractname}{Abstract.}
\renewcommand{\contentsname}{Table of Contents}
\renewcommand{\figurename}{Fig.}
\renewcommand{\tablename}{Table}

\newdimen\headlineindent             
\def\ps@headings{\let\@mkboth\@gobbletwo
   \let\@oddfoot\@empty\let\@evenfoot\@empty
   \def\@evenhead{\normalfont\small\rlap{\thepage}\hspace{\headlineindent}%
                  \leftmark\hfil}
   \def\@oddhead{\normalfont\small\hfil\rightmark\hspace{\headlineindent}%
                 \llap{\thepage}}
   \def\chaptermark##1{}%
   \def\sectionmark##1{}%
   \def\subsectionmark##1{}}

\def\ps@titlepage{\let\@mkboth\@gobbletwo
   \let\@oddfoot\@empty\let\@evenfoot\@empty
   \def\@evenhead{\normalfont\small\rlap{\thepage}\hspace{\headlineindent}%
                  \hfil}
   \def\@oddhead{\normalfont\small\hfil\hspace{\headlineindent}%
                 \llap{\thepage}}
   \def\chaptermark##1{}%
   \def\sectionmark##1{}%
   \def\subsectionmark##1{}}

\if@runhead\ps@headings\else
\ps@empty\fi

\makeatother

\newcommand{\smallspace}{\mskip 2mu minus 1mu}
\newcommand{\tinyspace}{\mskip 1mu}

\def\squareforqed{\hbox{\rlap{$\sqcap$}$\sqcup$}}
\def\qed{\ifmmode\squareforqed\else{\unskip\nobreak\hfil
\penalty50\hskip1em\null\nobreak\hfil\squareforqed
\parfillskip=0pt\finalhyphendemerits=0\endgraf}\fi}

\newcommand{\ket}[1]{\mbox{$| #1 \rangle$}}
\newcommand{\bra}[1]{\mbox{$\langle #1 |$}}
\newcommand{\inner}[2]{\mbox{$\langle #1 | #2 \rangle$}}

\newcommand{\nth}[1]{$#1$th}
\newcommand{\emm}{m}

\newcommand{\reals}{\mathbb{R}}

\newcommand{\weight}{\ensuremath{\omega}}

\newcommand{\op}[1]{{\mathsf{#1}}}

\newcommand{\projection}{\op{P}}

\newcommand{\color}{{\mathcal C}}
\newcommand{\tree}{{\mathcal T}}
\newcommand{\pathfx}{{\mathcal P}}
\newcommand{\leaf}[1]{\ell_{#1}}
\newcommand{\gleaf}{\ell}
\newcommand{\leftst}{\text{left}}
\newcommand{\rightst}{\text{right}}
\newcommand{\leaves}{{\mathcal L}}

\newcommand{\snorm}[1]{\ensuremath{|\!|\!|{#1}|\!|\!|_2}}
\newcommand{\bstr}[1]{\ensuremath{\{0, 1\}^{#1}}}
\newcommand{\lb}[1]{\ensuremath{\Omega({#1})}}

\newcommand{\component}[2]{\mbox{$\ket{{#1}\!\!\upharpoonright_{#2}}$}}

\makeatletter
\def\mylabel#1#2#3{\@bsphack
  \protected@write\@auxout{}%
         {\string\newlabel{#1}{{#2}{#3}}}%
  \@esphack}
\makeatother

\begin{document}


\mainmatter

\title{Quantum Complexities of Ordered Searching,\\ 
Sorting, and Element Distinctness%
\thanks{Research supported by 
the EU fifth framework program QAIP, IST-1999-11234,
and the National Science Foundation under grant CCR-9820855.}}

\titlerunning{\ }

\author{Peter H{\o}yer\,\inst{1}\fnmsep\thanks{Research conducted 
in part while at BRICS, University of Aarhus, Denmark.}
\and Jan Neerbek\,\inst{2}
\and Yaoyun Shi\,\inst{3}}

\authorrunning{}

\institute{Dept.\ of Comp.\ Sci., 
University of Calgary, Alberta, Canada T2N~1N4\\
\email{{hoyer}\boldmath$\mathchar"40$\texttt{cpsc.ucalgary.ca}}
\and
Dept.\ of Comp.\ Sci., University of Aarhus, 
\mbox{DK--8000} {\AA}rhus~C, Denmark\\
\email{{neerbek}\boldmath$\mathchar"40$\texttt{daimi.au.dk}}
\and 
Dept.\ of Comp.\ Sci., Princeton University, Princeton, 
\mbox{NJ~08544}, USA\\
\email{{shiyy}\boldmath$\mathchar"40$\texttt{cs.princeton.edu}}}



\maketitle

\begin{abstract}
We~consider the quantum complexities of the
following three problems:
searching an ordered list, sorting an un-ordered list,
and deciding whether the numbers in a list are all distinct.
Letting $N$ be the number of elements in the input list, we prove 
a lower bound of $\frac{1}{\pi}(\ln(N)-1)$ 
accesses to the list elements for ordered searching, 
a lower bound of $\Omega(N\log{N})$
binary comparisons for sorting,
and a lower bound of $\Omega(\sqrt{N}\log{N})$ binary comparisons
for element distinctness. The previously best known lower bounds
are $\frac{1}{12}\log_2(N) - O(1)$ due to Ambainis, $\Omega(N)$,
and $\Omega(\sqrt{N})$, respectively.
Our proofs are based on a weighted all-pairs inner product argument.

\parindent 15pt In addition to our lower bound results, 
we give a quantum algorithm 
for ordered searching using 
roughly $0.631 \log_2(N)$ oracle accesses.
Our algorithm uses a quantum routine for traversing 
through a binary search tree faster than classically,
and it is of a nature very different {from} a faster algorithm due to
Farhi, Goldstone, Gutmann, and Sipser. 
\end{abstract}


\section{Introduction}
\label{sec:intro}
The speedups of quantum algorithms over classical algorithms
have been a main reason for the current interests on
quantum computing. One central question regarding the power
of quantum computing is: How much speedup is possible?
Although dramatic speedups seem possible, as in the case of
Shor's~\cite{Shor} algorithms for factoring and 
for finding discrete logarithms,  
\emph{provable} speedups are found only
in restricted models such as the black~box model.

In the black~box model, the input is given as a black~box, so that the
only way the algorithm can obtain information about the input is via
queries, and the complexity measure is the number of queries.
Many problems that allow provable quantum speedups can be formulated
in this model, an example being the unordered search problem
considered by Grover~\cite{Grover}.  Several tight lower bounds 
are now known for this model, most of them being based on
techniques introduced in~\cite{BBBV,BBCMdW,Ambainis2}.

We~study the quantum complexities of the following three problems.
\begin{description}
\item[Ordered searching] Given a list of numbers 
$x = (x_0, x_1, \ldots, x_{N-1})$ 
in non-decreasing order and some number~$y$, 
find the minimal $i$ such that $y \le x_i$. 
We~assume that $x_{N-1} = \infty > y$ so that the problem is
always well-defined.
\item[Sorting] Given a list of numbers
$x = (x_0, x_1, \ldots, x_{N-1})$,
output a permutation $\sigma$ on the set $\{0,\ldots,N-1\}$
so that the list $(x_{\sigma(0)}, x_{\sigma(1)}, \ldots, x_{\sigma(N-1)})$
is in non-decreasing order. 
\item[Element distinctness] Given a list of numbers
$x = (x_0, x_1, \ldots, x_{N-1})$, are they all distinct?
\end{description}
These problems are closely related and 
are among the most fundamental and most studied 
problems in the theory of algorithms.
They can also be formulated naturally in the black~box 
model.
For the ordered searching
problem, we consider queries of the type 
\mbox{``$x_i = \tinyspace ?$''},
and for the sorting and element distinctness 
problems, we consider queries of the
type \mbox{``Is $x_i < x_{i'}$?''}, which are simply
binary comparisons.
Let $H_i = \sum_{k=1}^{i} \frac{1}{k}$ denote
the \nth{i} harmonic number.
We~prove a lower bound for each of these three problems.
\begin{theorem}\label{thm:search}
Any quantum algorithm for ordered searching that errs with
probability at most~$\epsilon \geq 0$ requires at least
\begin{equation}
\bigg(1-2\sqrt{\epsilon(1-\epsilon)}\bigg) 
\frac{1}{\pi}\tinyspace(H_N-1)
\end{equation}
queries to the oracle.
In~particular, any exact quantum algorithm requires more than
$\frac{1}{\pi}(\ln(N)-1) \approx 0.220\log_2{N}$ queries.
\end{theorem}
\begin{theorem}\label{thm:sort}
Any comparison-based quantum algorithm for sorting that errs with
probability at most~$\epsilon \geq 0$ requires at least
\begin{equation}
\bigg(1-2\sqrt{\epsilon(1-\epsilon)}\bigg) 
\frac{N}{2\pi}\tinyspace (H_N-1)
\end{equation}
comparisons.
In~particular, any exact quantum algorithm requires more than
$\frac{N}{2\pi}(\ln(N)-1) \approx 0.110N \log_2{N}$  comparisons.
\end{theorem}
\begin{theorem}\label{thm:ed}
Any comparison-based quantum algorithm for element distinctness
that errs with
probability at most~$\epsilon \geq 0$ requires at least
\begin{equation}
\bigg(1-2\sqrt{\epsilon(1-\epsilon)}\bigg) 
\frac{\sqrt{N}}{2\pi}\tinyspace (H_N-1)
\end{equation}
comparisons.
\end{theorem}

The previously best known quantum lower bound 
for ordered searching is $\frac{1}{12} \log_2(N) - O(1)$,
due to Ambainis~\cite{Ambainis1}. 
For comparison-based sorting and element distinctness, 
the previously best known quantum lower bounds are respectively 
$\Omega(N)$ and~$\lb{\sqrt{N}}$, 
both of which can be proven in many ways.

We~prove our lower bounds by utilizing what we refer to as 
a weighted all-pairs inner product argument,  
or a probabilistic adversary argument.
This proof technique is based on the work of 
Bennett, Bernstein, Brassard, and Vazirani~\cite{BBBV}
and Ambainis~\cite{Ambainis2}.

Farhi, Goldstone, Gutmann, and Sipser~\cite{FGGS2} have 
given an exact quantum algorithm for ordered searching using
roughly $0.526 \log_2(N)$ queries. 
We~provide an alternative quantum algorithm that 
is exact and uses $\log_3(N) + O(1) \approx 0.631 \log_2(N)$ queries.
Our construction is radically different {from} the construction
proposed by Farhi \emph{et~al.}~\cite{FGGS2}, and 
these are the only constructions known leading to
quantum algorithms using at most $c \log_2(N)$ queries 
for some constant~$c$ strictly less than~1.

Whereas most quantum algorithms are based on Fourier transforms and
amplitude amplification~\cite{BHMT}, our algorithm is based on binary
search trees.
We~initiate several applications of the binary search algorithm
in quantum parallel and let them find the element we are
searching for in teamwork.
By~cooperating, these applications can traverse the binary
search tree faster than classically,
hereby reducing the complexity {from} $\log_2(N)$ to
roughly $\log_3(N)$.

There are at least three reasons why the quantum complexities
of the three problems are of interest.
Firstly because of their significance in algorithmics in general.
Secondly because these problems possess some symmetries and 
periodicities of a different nature than other studied problems
in quantum algorithmics.
Determining symmetries and periodicities seems to be a primary
ability of quantum computers and it is not at all clear 
how far-reaching this skill~is.
Thirdly because searching and sorting represent
non-Boolean non-symmetric functions.
A~(partial) function is said to be symmetric if
it is invariant under permutation of its input.
Only few non-trivial quantum bounds for non-Boolean 
and non-symmetric functions are known.

The rest of the paper is organized as follows.
We~first discuss the model in Sect.~\ref{sec:blackbox},
present our general technique for proving lower bounds in
Sect.~\ref{subsec:general}, and then apply it to the three problems
in Sects.~\mbox{\ref{subsec:search}--\ref{subsec:elem}}.
We~give our quantum algorithm for ordered searching 
in Sect.~\ref{sec:upper} and conclude in Sect.~\ref{sec:conclusion}.

\section{Quantum black~box computing}
\label{sec:blackbox}
We give a formal definition of the black~box model, 
which is slightly different {from}, but equivalent to, 
the definition of 
Beals, Buhrman, Cleve, Mosca, and de~Wolf given in~\cite{BBCMdW}.
Fix some positive integer $N>0$.  
The input $x =(x_0,\ldots,x_{N-1})\in \{0,1\}^N$ 
is given as an oracle, and the only way we can access the 
bits of the oracle is via queries.  
A~query implements the operator
\begin{equation}\label{eq:oracle}
\op{O}_x: \quad \ket{z;i} \;\longmapsto\; 
\begin{cases}
(-1)^{x_i} \ket{z;i} & \text{if $0\leq i < N$}\\
\hphantom{(-1)^{x_i}} \ket{z;i} & \text{if $i \geq N$.}
\end{cases}
\end{equation}
Here $i$ and $z$ are non-negative integers.
By~a query to oracle~$x$ we mean an application of the 
unitary operator~$\op{O}_x$.  
We~sometimes refer to~$\op{O}_x$ as the oracle.
A~quantum algorithm~$\op{A}$ that uses $T$ queries to an 
oracle~$\op{O}$ is a unitary operator of the form
\begin{equation}\label{eq:algorithm}
\op{A} = (\op{U} \op{O})^T \op{U}.
\end{equation}
We~always apply algorithm~$\op{A}$ on the initial state~$\ket{0}$.
For every integer $j \geq 0$ and every oracle~$x$, let
\begin{equation}
\ket{\psi_x^j} \smallspace=\smallspace (\op{U}\op{O}_x)^j \op{U}
\smallspace\ket{0}
\end{equation}
denote the state after $j$ queries, given oracle~$x$.
After applying~$\op{A}$, we always measure the final 
state in the computational basis.

Consider the computation of some function 
$f: S \rightarrow \{0,1\}^\emm$, where $S \subseteq \{0,1\}^N$.  
We~say that
algorithm~$\op{A}$ computes $f$ with error probability bounded
by~$\epsilon$, for some constant~$\epsilon$ with $0\le \epsilon < 1/2$, 
if for any $x \in S$,
the probability of observing $f(x)$ when the $\emm$
rightmost bits of $\ket{\psi_x^T}$ are measured is at 
least \mbox{$1-\epsilon$}.

\section{Lower bounds}
\label{sec:lowerbounds}

\subsection{General technique}
\label{subsec:general}
We use the notation of Sect.~\ref{sec:blackbox}. 
For any $\epsilon \geq 0$, 
let $\epsilon'= 2 \sqrt{\epsilon(1-\epsilon)}$.

The computation always starts in the same initial state~$\ket{0}$,
so for all oracles $x \in S$ we have
$\ket{\psi^0_x} = \ket{0}$. 
If~for two input oracles $x,y \in S$, 
the correct answers are different, i.e., if $f(x) \ne f(y)$,
then the corresponding final states 
$\ket{\psi^T_x}$ and $\ket{\psi^T_y}$ must be 
almost orthogonal.

\begin{lemma}\label{lm:epsilon}
For all oracles $x, y \in S$ so that $f(x) \ne f(y)$, 
$|\inner{\psi^T_x}{\psi^T_y}| \leq \epsilon'$.
\end{lemma}

Now consider a probability distribution over those pairs of
inputs $(x,y) \in S \times S$ for which $f(x) \ne f(y)$. 
For each integer $j\ge 0$, we use the
following quantity to quantify the \emph{average progress} of
the algorithm in distinguishing any two inputs 
after applying $(\op{U}\op{O})^j\op{U}$,
\begin{equation*}
W_j \smallspace=\smallspace
 \mathbf{E}_{(x, y)} \left[\;\inner{\psi_x^j}{\psi_y^j}\; \right].
\end{equation*}
Observe that $W_0 = 1$ 
and that $W_T \le \epsilon'$ by Lemma~\ref{lm:epsilon}. 
By~proving that for every~$j$ with $0 \le j < T$, we have
$|W_{j} - W_{j+1}| \le \delta$, we conclude that
$T \ge (1-\epsilon')/\delta$.

For simplicity of presentation, we scale the probabilities by 
using a \emph{weight function} 
$\weight: \, S \times S \rightarrow \reals^{+}$.
{From} now on, we use the following definition 
of~$W_j$ to quantify the overall progress of the algorithm,
\begin{equation}
\label{eq:progress}
W_j =
 \sum_{x,y \in S} \weight(x,y)\; \inner{\psi_x^j}{\psi_y^j}.
\end{equation} 

Our technique is a natural generalization of 
Ambainis' approach \cite{Ambainis2}, which uses
uniform distributions over subsets of $S \times S$.
Our lower bound proofs imply that non-uniform
distributions can give better lower bounds. 
Clearly, finding a ``good'' distribution is an important 
step in applying our technique.
Another important step is to find a tight bound on the 
progress after each oracle query.

We~end this subsection by introducing some notation
and stating two lemmas we require when bounding the progress.
For every $i\geq 0$, let
$\projection_i = \sum_{z\geq 0} \ket{z;i}\bra{z;i}$
denote the projection operator onto the subspace
querying the \nth{i} oracle bit. For $i < 0$,
operator $\projection_i$ is taken as the zero projection.
The following lemma, which may be proven by
the Cauchy--Schwarz inequality, bounds the quantified 
progress that one oracle query makes
in distinguishing two inputs $x$ and~$y$.

\begin{lemma}
\label{lm:progress}
For any oracles $x, y \in \bstr{N}$, and any integer $j \ge 0$,
\begin{equation}
|\inner{\psi^{j}_x}{\psi^{j}_y}
- \inner{\psi^{j+1}_x}{\psi^{j+1}_y}| \;\leq\;  2
\sum_{i: x_i \ne y_i} \|\projection_i\ket{\psi^j_x}\| \cdot
\|\projection_i\ket{\psi^j_y}\|.
\end{equation}
\end{lemma}
We~sometimes write $\ket{\psi_x}$ as shorthand for $\ket{\psi^j_x}$
once integer $j$ is fixed.

Let $A = [\alpha_{k, \ell}]_{1\le k, \ell<\infty}$
be the Hilbert matrix with
$\alpha_{k, \ell} = 1/(k+\ell-1)$, and
$\snorm{\cdot}$ be the spectral norm, i.e.,
for any complex-valued matrix $M \in \mathbb{C}^{m \times m}$,
the norm $\snorm{M}$ is defined as
$\max \{\|Mx\|_2 \} $, where the maximum is taken over
all unit vectors $x \in \mathbb{C}^m$.
Let $B_N = [\beta_{k, \ell}]_{1\le{}k, \ell\le N}$ 
be the matrix where entry $\beta_{k,\ell}$ is 
$\frac{1}{k+\ell-1}$ if $k+\ell\le N+1$, and $0$ otherwise.
Clearly $\snorm{B_N} \le \snorm{A}$ for any $N>0$.
Our lower bound proofs
rely on the following
property of the Hilbert matrix.
\begin{lemma}[E.g.: Choi~\cite{Choi}] \label{lm:choi}
$\snorm{A} = \pi$. 
Hence, $\snorm{B_N} \le \pi$.
\end{lemma}

\subsection{Lower bound for ordered searching}
\label{subsec:search}
The first non-trivial quantum lower bound on ordered searching
proven was $\Omega(\sqrt{\log_2(N)}/\log_2\log_2(N))$, due to
Buhrman and de~Wolf~\cite{BdW}
by an ingenious reduction {from} the \textsc{parity} problem.
Farhi, Goldstone, Gutmann, and Sipser~\cite{FGGS1}
improved this to $\log_2(N)/2\log_2\log_2(N)$, and
Ambainis~\cite{Ambainis1} then proved the
previously best known lower bound of $\frac{1}{12} \log_2(N) - O(1)$.
In~\cite{FGGS1,Ambainis1}, they use, as we do here,
an inner product argument along the lines of~\cite{BBBV}.
In~this section, we improve the lower bound by a constant factor.

For the purpose of proving the lower bound, we assume that 
each of the $N$ input numbers is either $0$ or~$1$,
and that the input does not consist of all zeroes.
That is, the set $S$ of possible inputs are the ordered $N$-bit strings
of non-zero Hamming weight.
The search function $f:S \rightarrow \{0,1\}^\emm$ 
is defined by $f(x) = \min\{ 0 \leq i<N \mid x_i =1\}$,
where we identify the result $f(x)$ 
with its binary encoding as a bit-string of 
length $\emm = \lceil \log_2(N)\rceil$.
As~our weight function~$\weight$, we choose the inverse of the 
difference in Hamming weights,
\begin{equation}\label{eq:weight}
\weight(x,y) =
\begin{cases}
\frac{1}{f(y)-f(x)}
 &\text{if $0 \le f(x) < f(y) < N$}\\
0  &\text{otherwise.}
\end{cases}
\end{equation}
With this choice, we have that $W_0 = N H_N - N$
and by Lemma~\ref{lm:epsilon} also that
$W_T \leq \epsilon' W_0$.
Theorem~\ref{thm:search} then follows {from} the next lemma.

\begin{lemma}\label{lm:search}
For every $j$ with $0 \leq j <T$ we have that
$|W_{j}-W_{j+1}| \leq \pi N$.
\end{lemma}

\begin{proof}
As~shorthand, we write $\ket{\psi_{f(x)}}$ for $\ket{\psi^j_x}$.
By Lemma~\ref{lm:progress},
\begin{align*}
|W_{j} - W_{j+1}| 
&\;\le\; 2\smallspace
   \sum_{k=0}^{N-2} \sum_{\ell=k+1}^{N-1} \frac{1}{\ell-k} \sum_{i=k}^{\ell-1}
   \|\projection_i\ket{{\psi_k}}\| \cdot 
   \|\projection_i\ket{{\psi_\ell}}\|\\
&\;=\; 2\smallspace 
   \sum_{d=1}^{N-1} \smallspace \sum_{i=0}^{d-1} \tinyspace
   \frac{1}{d} \tinyspace \sum_{k=0}^{N-d-1}
       \|\projection_{k+i}\ket{{\psi_k}}\| \cdot
   \|\projection_{k+i}\ket{\psi_{k+d}}\|.
\end{align*}
Let vectors $\gamma = [\gamma_i]_{0 \leq i < N-1} \in \mathbb{R}^{N-1}$
and $\delta = [\delta_i]_{0 \leq i < N-1} \in \mathbb{R}^{N-1}$
be defined~by
\begin{equation*}
\gamma_i = \Bigg(\sum_{k=0}^{N-1} 
  \|\projection_{k+i} \ket{\psi_k}\|^2\Bigg)^{1/2}
\;\quad \text{and }\quad \;\quad
\delta_i = \Bigg(\sum_{k=0}^{N-1} 
  \|\projection_{k-i-1} \ket{\psi_k}\|^2\Bigg)^{1/2}.
\end{equation*}
Then, by the Cauchy--Schwarz inequality,
\begin{equation}
\label{eq:matrixproduct}
|W_{j} - W_{j+1}|
\leq 2 \sum_{d=1}^{N-1} \sum_{i=0}^{d-1} \frac{1}{d} 
  \gamma_i \delta_{d-i-1} \;=\; 2 \gamma^t B_N \delta,
\end{equation}
where $t$ denotes matrix transposition.
Since each vector $\ket{\psi_k}$ is of unit norm, we have 
$\|\gamma\|_2^2 + \|\delta\|_2^2 \leq N$, so 
$\|\gamma\|_2  \|\delta\|_2 \leq N/2$.
The matrix product $2 \gamma^t B_N \delta$ is upper bounded~by
$2 \|\gamma\|_2 \cdot \snorm{B_N} \cdot \|\delta\|_2$,
which is at most $\pi N$ by Lemma~\ref{lm:choi}.
\qed
\end{proof}

\subsection{Lower bound for sorting}
\label{subsec:sort}
We assume that the $N$ numbers to be sorted, $x = (x_0, \ldots, x_{N-1})$,
correspond to some permutation $\sigma$ on $\{0, 1, \ldots, N-1\}$.
That is, $x_i = \sigma(i)$ for every $0 \leq i < N$.
We~assume the input to the quantum algorithm is 
the comparison matrix $M_\sigma = [m_{ii'}]_{0\le i, i' < N}$ with
\begin{equation*}
m_{ii'} = 
\begin{cases}1 & \text{ if $\sigma(i) < \sigma(i')$}\\
0 & \text{ otherwise.}\end{cases}
\end{equation*}
One comparison corresponds to one application of the oracle operator
\begin{equation*}
\op{O}_\sigma = \sum_{z \geq 0}
     \sum_{i,i' \geq 0} (-1)^{m_{ii'}}\ket{z;i,i'}\bra{z;i,i'}.
\end{equation*}
To~simplify notation, we sometimes identify the 
input~$M_\sigma$ with the underlining permutation~$\sigma$.

For every pair $\{i,i'\}$ of indices with $0 \leq i, i' <N$, let
\[\projection_{ii'} = \sum_{z\geq 0} \ket{z;i,i'}\bra{z;i,i'}
  + \sum_{z\geq 0} \ket{z;i',i}\bra{z;i',i}\]
denote the projection operator onto the subspace comparing
the \nth{i} and \nth{(i')} elements.
For any vector $\ket{\psi}$, 
we use $\component{\psi}{\sigma, k, \ell}$ as shorthand for
$\projection_{\sigma^{-1}(k), \sigma^{-1}(\ell)} \ket{\psi}$.

For every permutation~$\sigma$,
and every integers $0 \leq k \leq N-2$ and $1 \leq d \leq N-1-k$,
define a new permutation,
\begin{equation}
\label{eq:sigmakd}
\sigma^{(k,d)} = (k, k+1, \ldots, k+d) \circ \sigma.
\end{equation}
If $\tau = \sigma^{(k,d)}$, then
\begin{equation}\label{eq:inverse}
\sigma^{-1}(i) = \begin{cases}
\tau^{-1}(k) & \text{ if $i=k+d$}\\
\tau^{-1}(i+1) & \text{ if $k \le i < k+d$}\\
\tau^{-1}(i) & \text{ otherwise.}
\end{cases}
\end{equation}
This implies that the comparison matrices 
$M_\sigma$ and $M_\tau$ differ only on the following 
pairs of entries,
\begin{equation}
\label{eq:diff}
\big\{\sigma^{-1}(k+d), \sigma^{-1}(k+i)\big\}
= \big\{\tau^{-1}(k), \tau^{-1}(k+i+1)\big\}
\end{equation}
for all $i$ with $0 \leq i < d$.

Informally, if $M_\sigma$ corresponds to some list~$x$, 
then $M_\tau$ corresponds to the list~$y$ obtained by 
replacing the element of rank~$k+d$ in~$x$ by a new element 
of rank~$k$ (the element in $x$ that had rank $k$ then 
has rank $k+1$ in $y$, etc.).
The only way the algorithm can distinguish $\sigma$ {from} $\tau$ 
is by comparing the element of rank $k+d$ in~$x$ with
one of the $d$ elements of rank $k+i$ for some $0 \leq i < d$.

We~choose the following weight function,
\begin{equation}\label{eq:weightsort}
\weight(\sigma, \tau) =
\begin{cases}
\frac{1}{d}
 &\text{if $\tau = \sigma^{(k,d)}$ for some $k$ and $d$}\\
0  &\text{otherwise.}
\end{cases}
\end{equation}
Then one may verify that 
$W_0 = N!\smallspace (N H_N - N)$, and $W_T \le \epsilon' W_0$.
To prove Theorem~\ref{thm:sort}, we need only to prove the following lemma.
\begin{lemma}\label{lm:sort}
For any $j$ with $0 \leq j <T$,
$|W_{j}-W_{j+1}| \leq 2\pi \tinyspace N!$. 
\end{lemma}

\begin{proof} 
Similar to the proof of Lemma~\ref{lm:search}.
By Lemma~\ref{lm:progress} and~(\ref{eq:diff}), 
\begin{equation*}
|W_{j} - W_{j+1}| \leq 
 2 \sum_{d=1}^{N-1} \sum_{i=0}^{d-1} \frac{1}{d}
    \sum_{\sigma} \sum_{k=0}^{N-d-1} \;
    \|\component{\psi_\sigma}{\sigma, k+d, k+i}\| \cdot
    \|\component{\psi_{\sigma^{(k,d)}}}{\sigma, k+d, k+i}\|.
\end{equation*}
Let $\gamma = [\gamma_i]_{1\le{}i<N} \in \mathbb{R}^{N-1}$ 
be such that
$\gamma_i = \big( {
       \sum_{\sigma} \sum_{\ell=0}^{N-1}
        \big\|
       \component{\psi_\sigma}{\sigma, \ell, \ell+i}
       \big\|^2 }\big)^{1/2}$,
where we let $\ell$ range {from} $0$ to~$N-1$ and simply set
the thus caused undefined projection operators to be zero
operators.
Then by~(\ref{eq:inverse}), 
\begin{equation*}
\sum_{\sigma} \sum_{k=0}^{N-d-1}
\big\|
\component{\psi_{\sigma^{(k,d)}}}{\sigma, k+d, k+i} 
\big\|^2 \;=\;
\sum_{\tau} \sum_{k=0}^{N-d-1}
\big\|
\component{\psi_\tau}{\tau, k, k+i+1}
\big\|^2 \;\le\; \gamma_{i+1}^2.
\end{equation*}

Applying the Cauchy--Schwarz inequality,
and in analogy with~(\ref{eq:matrixproduct}),
\begin{equation}
\label{eq:matrixprod}
|W_{j} - W_{j+1}| 
\leq 2 \sum_{d=1}^{N-1} \sum_{i=0}^{d-1} \frac{1}{d}
  \gamma_{d-i} \gamma_{i+1}
\;=\; 2 \gamma^t B_{N-1} \gamma.
\end{equation}
Since $\|\gamma\|_2^2 \leq N!$, we conclude 
that $|W_j-W_{j+1}| \leq 2\pi \tinyspace N!\tinyspace$.
\qed
\end{proof}

\subsection{Lower bound for element distinctness}
\label{subsec:elem}
We modify the adversary for sorting as follows. 
As~in Sect.~\ref{subsec:sort}, when we talk about permutations,
the underlying set is $\{0, 1, \ldots, N-1\}$.
\begin{definition}
An annotated permutation is a permutation~$\tau$ with a marker 
on a single element $r_\tau$ for some $0\le r_\tau<N-1$.
\end{definition}
For every permutation $\sigma$, and every integers $k$ and $d$ as in
Sect.~\ref{subsec:sort}, the annotated
permutation $\tau = \sigma^{(k,d)}$ is the
same permutation as in~(\ref{eq:sigmakd}) but with
the rank $k$ element marked.
The only places where $M_\sigma$ and 
$M_\tau$ differ, are at the same entries as those 
in~(\ref{eq:diff}).

We use the same weight function as in~(\ref{eq:weightsort}).
Then $W_0 = N!\smallspace (N H_N - N)$ and $W_T \le \epsilon' W_0$.
We need only to prove the following lemma.

\begin{lemma}\label{lm:ed}
For any integer $j$ with $0 \leq j <T$,
$|W_j-W_{j+1}| \leq 2 \pi N! \sqrt{N}$.
\end{lemma}

\begin{proof} Almost identical to
the proof for Lemma~\ref{lm:sort},
except that we now require a second vector
$\delta = [\delta_{i}]_{1\le{}i<N}\in\mathbb{R}^{N-1}$ 
with
$\delta_i = \big( {
       \sum_{\tau} 
        \big\|
       \component{\psi_\tau}{\tau, r_\tau, r_\tau + i}
       \big\|^2 }\big)^{1/2}$.
Then by~(\ref{eq:inverse}),
\begin{equation*}
\sum_{\sigma} \sum_{k=0}^{N-d-1}
\big\|
\component{\psi_{\sigma^{(k,d)}}}{\sigma, k+d, k+i}
\big\|^2 \;=\;
\sum_{\tau: r_\tau<N-d}
\big\|
\component{\psi_\tau}{\tau, r_\tau, r_\tau+i+1}
\big\|^2 \;\le\; \delta_{i+1}^2.
\end{equation*}
In analogy with~(\ref{eq:matrixprod}), we have
\begin{equation*}
|W_{j} - W_{j+1}|
\leq 2 \sum_{d=1}^{N-1} \sum_{i=0}^{d-1} \frac{1}{d}
  \gamma_{d-i} \delta_{i+1}
\;=\; 2 \gamma^t B_{N-1} \delta.
\end{equation*}
Besides having $\|\gamma\|^2 \le N!$ as in the proof
of Lemma~\ref{lm:sort}, we also have that
\begin{equation*}
\|\delta\|^2 = 
   \sum_{i=1}^{N-1} 
   \sum_\tau \big\|
       \component{\psi_\tau}{\tau, r_\tau, r_\tau + i}
       \big\|^2  \le N! (N-1) \le N! N.
\end{equation*}
Therefore, $|W_{j} - W_{j+1}| \le 2 \pi \sqrt{N!} \sqrt{N!N} = 
2 \pi N! \sqrt{N}$.
\qed
\end{proof}

\section{A~{{\boldmath$\log_3(N)$}} algorithm for ordered searching}
\label{sec:upper}

We~begin by considering binary search trees on which our 
quantum algorithm is based. 
Let $\tree$ be a binary tree with $N \geq 2$ leaves.
We~put colored pebbles on the (internal) vertices of~$\tree$ 
subject to the following 2~conditions:
\begin{enumerate}
\item[(A)]
on every path {from} the root of $\tree$ to a leaf,
there is exactly 1 pebble of each color, and
\item[(B)] the number of pebbles $p_v$ on any vertex~$v \in \tree$
is at least as large as the total number of pebbles on 
its proper ancestors.
\end{enumerate}
We~say that $\tree$ is \emph{covered by $N'$ pebbles} if 
we can satisfy the 2~above conditions using at most $N'$ pebbles 
of each color.
We~want to minimize the maximum number~$N'$ of pebbles used of any color. 
We~say a covering is \emph{fair} if it uses the same number 
of pebbles of every color.
We~say a covering is \emph{tight} if, for all vertices $v \in \tree$, 
we have that $p_v$ equals the total number of pebbles on 
its proper ancestors, or there are no pebbles 
on any of the ancestors of~$v$.  
We~require the following two lemmas.

\begin{lemma}\label{lm:cover}
For every even integer $N\geq 2$, 
there exists a binary tree with $N$ leaves 
that can be fairly and tightly covered by 
$N' = \lfloor\frac{1}{3}N + \log_2(N)\rfloor$ pebbles
using $2^s$ colors, where $s=\lfloor \log_4(N/2)\rfloor$.
\end{lemma}

\begin{lemma}\label{lm:recursion}
Let integer-valued function $\tilde F$ be recursively defined by
\begin{equation*}
\tilde F(N) = 
\begin{cases}
\tilde F\big(\lfloor\frac{1}{3}N + \log_2(N)+1\rfloor\big) +1 
& \text{ if $N >8$}\\
1 & \text{ if $N \leq 8$.}
\end{cases}
\end{equation*}
Then $\tilde F(N) = \log_3(N) + O(1)$.
\end{lemma}

As~in Sect.~\ref{subsec:search}, we assume 
the oracle $x=(x_0,\ldots,x_{N-1}) \in \{0,1\}^N$ is a binary
string of non-zero Hamming weight.
The problem is to determine the leftmost~1 in~$x$, 
that is, to compute $f(x) = \min\{ 0 \leq i<N \mid x_i =1\}$.
Let $\tree$ be a binary tree with $N$ leaves for which
Lemma~\ref{lm:cover} holds.
Let $s=\lfloor \log_4(N/2)\rfloor$ and
$N' = \lfloor\frac{1}{3}N + \log_2(N)\rfloor$ be 
as in the lemma.
We~label the $N$ leaves of~$\tree$ by $\{0,\ldots,N-1\}$ {from} left 
to right. 
Let $\leaf{f(x)}$ denote the leaf labelled by~$f(x)$, and
let $\pathfx$ denote the path {from} the root of $\tree$ to
the parent of~$\leaf{f(x)}$.
We~think of $\pathfx$ as the path the classical search 
algorithm would traverse if searching for~$f(x)$ in tree~$\tree$.

Let $\color =\{c_0,\ldots,c_{2^{s}-1}\}$ be the set of $2^s$ colors
used in Lemma~\ref{lm:cover}.
For each color $c \in \color$,
let $V_c$ denote the set of vertices in $\tree$ 
populated by a pebble of color~$c$.
By~Condition~(A), there are at most $N'$ such vertices, that is,
$|V_c| \leq N'$.
Let $v_c$ denote the unique vertex in $V_c$ that is 
on path~$\pathfx$.
We~think of vertex $v_c$ as the root of the subtree ``containing''
leaf~$\leaf{f(x)}$.
Note that, by definition,
$v_c \in \pathfx$ for every color $c \in \color$, and that
$\sum_{v \in \pathfx} p_v = 2^s$ by \mbox{Condition~(A)}.

Our algorithm utilizes 3 unitary operators, 
$\op{U}_1$, $\op{O}_{x}'$, and $\op{U}_2$.
The first operator, $\op{U}_1$, is defined~by
\begin{equation}\label{eq:U1}
\op{U}_1: \quad
\ket{v}\ket{0} \;\longmapsto\; 
\ket{v} \bigg(\frac{1}{\sqrt{p_v}} 
\sum_c \ket{c}\bigg)
\qquad (v \in \tree),
\end{equation}
where the summation is over all colors $c \in \color$ that
are represented by a pebble on vertex~$v$.
We~refer to $\op{U}_1$ as the \emph{coloring operator}
and its inverse as the \emph{un-coloring operator}.

The query operator $\op{O}_{x}'$ is defined~by
\begin{equation}\label{eq:Ox}
\op{O}_{x}': \;\;
\ket{v} \,\longmapsto\,
\left\{
\begin{split}
\ket{v\smallspace;\tinyspace x_i}
   \quad&\text{ if there are no pebbles on the parent of $v$}\\
(-1)^{x_i} \ket{v}
   \quad&\text{ otherwise,} 
\end{split}
\right.
\end{equation}
where $i$ denotes the label of the rightmost leaf in the left
subtree of vertex~$v$.
Query operator~$\op{O}_{x}'$ is clearly unitary 
(or rather, can be extended to a unitary operator since 
it is only defined on a proper subspace).
Operator~$\op{O}_{x}'$ is slightly different {from}, but equivalent to, 
the query operator defined in Sect.~\ref{sec:blackbox}.
It~mimics the classical search algorithm by
querying the bit~$x_i$ that corresponds to the rightmost leaf in 
the left subtree of~$v$.

We~also use a unitary operator $\op{U}_2$ that maps 
each vertex to a superposition over the leaves in its subtree.
For every vertex and leaf $u$ in~$\tree$, let $\leaves(u)$ 
denote the set of leaves in the subtree rooted at~$u$,
and let
\begin{equation}\label{eq:phi}
\ket{\Phi_{u}} = 
\sum_{\gleaf \in \leaves(u)}
  \frac{1}{\sqrt{2^{d(u,\gleaf)}}} \ket{\gleaf},
\end{equation}
where $d(u,\gleaf)$ denotes the absolute value of the 
difference in depths of $u$ and leaf~$\gleaf$.
The unitary operator $\op{U}_2$ is (partially) defined as follows.
For every vertex $v \in \tree$ with no pebbles on its parent,
\stepcounter{equation}
\begin{align*}
\mylabel{eq:U2}{\theequation}{\thepage}
\ket{v\smallspace;\tinyspace 0} \;&\longmapsto\;
 \ket{\Phi_{\rightst(v)}}
\tag{\theequation .1}\\
\ket{v\smallspace;\tinyspace 1} \;&\longmapsto\;
 \ket{\Phi_{\leftst(v)}},
\tag{\theequation .2}
\intertext{and 
for every vertex $v \in \tree$ with pebbles on its parent,}
\ket{v} \;&\longmapsto\;
\frac{1}{\sqrt{2}}
  \big(\ket{\Phi_{\rightst(v)}} - \ket{\Phi_{\leftst(v)}}\big).
\tag{\theequation .3}
\end{align*}
Here $\leftst(v)$ denotes the left child of~$v$, and 
$\rightst(v)$ the right child.

Our quantum algorithm starts in the initial state $\ket{0}$ 
and produces the final state $\ket{\leaf{f(x)}}$.
Let $F(N)$ denote the number of queries used by the algorithm
on an oracle $x$ of size~$N$.
\begin{enumerate}

\item We first set up a superposition over all $2^s$ colors,
$\frac{1}{\sqrt{2^s}}\sum_{c \in \color} \ket{0}\ket{c}$.
\item We then apply our exact quantum search algorithm recursively.
For each color $c \in \color$ in quantum parallel, 
we search recursively among the vertices in~$V_c$, 
hereby determining the root $v_c \in V_c$ 
of the subtree containing the leaf~$\leaf{f(x)}$.
Since $|V_c| \leq N'$,
this requires at most $F(N'+1)$ queries to oracle~$x$ and 
produces the superposition
$\frac{1}{\sqrt{2^s}}\sum_{c \in \color} \ket{v_c}\ket{c}$.
Since every vertex $v_c$ in this sum is on the path $\pathfx$,
we can rewrite the sum~as
\begin{equation*}\label{eq:step2}
\frac{1}{\sqrt{2^s}}\sum_{v \in \pathfx} \ket{v}
\sum_{c \in \color : v_c=v} \ket{c}.
\end{equation*}

\item We then apply the un-coloring operator $\op{U}_1^{-1}$, 
producing the superposition
$\frac{1}{\sqrt{2^s}}\sum_{v \in \pathfx} \sqrt{p_v} \,\ket{v}\ket{0}$.
Ignoring the second register which always holds a~zero, this is
\begin{equation*}\label{eq:step3}
\frac{1}{\sqrt{2^s}} \sum_{v \in \pathfx} \sqrt{p_v} \,\ket{v}.
\end{equation*}
That is, we have (recursively) obtained a superposition
over the vertices on the path $\pathfx$ {from} the root of~$\tree$
to the parent of the leaf $\leaf{f(x)}$ labelled by~$f(x)$.
 
\item \label{item:haar}
We then apply the operator $\op{U}_2^{\vphantom{'}}\op{O}_{x}'$,
producing the final state
\begin{equation*}
\op{U}_2^{\vphantom{'}}\op{O}_{x}'
\frac{1}{\sqrt{2^s}} \sum_{v \in \pathfx} \sqrt{p_v} \,\ket{v}
\;=\;
\frac{1}{\sqrt{2^s}} \sum_{v \in \pathfx} \sqrt{p_v} 
\,\op{U}_2^{\vphantom{'}}\op{O}_{x}' \,\ket{v},
\end{equation*}
which one can show equal to~$\ket{\leaf{f(x)}}$.
Thus, a final measurement of this state yields~$f(x)$ with certainty.
\end{enumerate}

The total number of queries to the oracle~$x$ is
at most $F(N'+1)+1$, and thus, by 
Lemma~\ref{lm:recursion}, the algorithm uses at 
most $\log_3(N)+O(1)$ queries.  Theorem~\ref{thm:alg} follows.

\begin{theorem}\label{thm:alg}
The above described quantum algorithm for searching 
an ordered list of $N$ elements is exact and 
uses at most $\log_3(N)+O(1)$ queries.
\end{theorem}

\section{Concluding remarks and open problems}
\label{sec:conclusion}
The inner product of two quantum states is a measure for their 
distinguishability.  
We~have proposed a weighted all-pairs inner product 
argument as a tool for proving lower bounds in the 
quantum black~box model.
The possibility of using non-uniform weights seems 
particularly suitable when proving lower bounds 
for non-symmetric (possibly partial) functions.
It~could be interesting to consider other measures than inner
products, as discussed, for instance, by Zalka~\cite{Zalka},
Jozsa and Schlienz~\cite{JS}, and Vedral~\cite{Vedral}.

The result of 
Grigoriev, Karpinski, Meyer auf der Heide, and Smolensky~\cite{GKHS} 
implies that if only comparisons are allowed, 
the randomized decision tree complexity of element distinctness 
has the same $\Omega(N\log{N})$ lower bound as sorting. 
Interestingly, their quantum complexities differ dramatically: 
the quantum algorithm by Buhrman \emph{et~al.}~\cite{BDHHMSW} 
uses only $O(N^{3/4}\log{N})$ comparisons.  There is
still a big gap between this upper bound and 
our lower bound of $\Omega({N}^{1/2}\log{N})$.
One way of closing this gab might be to consider
quantum time-space tradeoffs, as has been 
done for the classical case~\cite{BFKLT,Beame}.

Our algorithm for searching an ordered list with complexity 
$\log_3(N) +O(1)$ is based on the classical binary search algorithm.
The quantum algorithm initiates several independent walks/searches
at the root of the binary search tree.
These searches traverse down the tree faster than classically by
cooperating, and they eventually all reach the leaf we are searching
for in roughly $\log_3(N)$ steps.
It could be interesting to consider if similar ideas 
can be used to speed up other classical algorithms.
For instance one may consider other applications 
of operators like~$\op{U}_2$ acting on rooted trees and graphs.


\section*{Acknowledgements}
We~are grateful to 
Andris Ambainis, Harry Buhrman, Mark Ettinger, 
Gudmund S.~Frandsen, Dieter van~Melkebeek, Hein R{\"o}hrig, 
Daniel Wang, Ronald de~Wolf, Andy~Yao, and especially Sanjeev~Arora, 
for their precious comments and suggestions.


Many of the above references can be found at the Los Alamos
National Laboratory e-print archive 
(http://arXiv.org/archive/quant-ph).

\end{document}